# Suspended Long-Lived NMR Echo in Solids

A. Turanov[1] and A.K. Khitrin[2]

[1] *Zavoisky Physical-Technical Institute RAS, Kazan, 420029, Russia*

[2] *Department of Chemistry, Kent State University, OH 44242, USA*

We report an observation of extremely long-lived spin states in systems of dipolar-coupled nuclear spins in solids. The "suspended echo" experiment uses a simple stimulated echo pulse sequence and creates non-equilibrium states which live many orders of magnitude longer than the characteristic time of spin-spin dynamics $T_2$. Large amounts of information can be encoded in such long-lived states, stored in a form of multi-spin correlations, and subsequently retrieved by an application of a single "reading" pulse.

Systems of coupled spins ½ play a special role in exploration of many-body quantum dynamics. Individual spins are the simplest two-level quantum objects, and all dynamical complexity comes from interactions between them. Nuclear spins are perfect experimental objects because their spin degrees of freedom can be extremely well isolated from the environment (lattice) and also because spin Hamiltonians are known with high accuracy. In addition, modern NMR spectrometers and existing experimental methods allow precise manipulation and high-sensitivity measurements. High-order multi-spin correlations can be directly measured in NMR experiments [1,2]. As another example of sophistication of available NMR techniques we can mention addressing individual quantum states in a cluster of 12 dipolar-coupled nuclear spins, a system with 4096 quantum levels [3]. The number of quantum levels $2^N$ grows very fast with the number of spins $N$. This restricts capabilities of direct simulations of spin dynamics by $N \approx 15$. At the same time, relative fluctuations decay only as $N^{-1/2}$. As a result, some phenomena, emerging at large $N$, and surviving in the thermodynamic limit $N \to \infty$, like the small-amplitude long-lived partial echo discussed in this Letter, are difficult to reproduce in first-principles simulations. For small clusters, manageable by simulations, the effects remain buried under fluctuations. So far, the experiments remain the only reliable guidance in studying such phenomena.

Long-lived states of small isolated clusters of nuclear spins and their applications is an interesting new development in liquid-state NMR. Such states, as an example, a singlet state of two equivalent spins, are immune to spin-lattice relaxation and can live much longer than the spin-lattice relaxation time $T_1$. The singlet state can be accessed by turning on and off the difference of resonance frequencies (chemical shifts) of the two spins [4]. Superpositions of the long-lived states, the long-lived coherences, can be used for improving spectral resolution [5]. Long-lived states for isolated clusters of three spins have been also identified [6]. Eigenstates of a spin Hamiltonian are stationary and can be individually addressed in small spin clusters. Therefore, in this case, the concept "long-lived state" can be related only to the strength of interaction with environment (spin-lattice relaxation).

The situation is very different in solids, where dipole-dipole interactions between nuclear spins are not averaged out by molecular motions, and all spins in a macroscopic sample can be coupled by a network of dipole-dipole interactions. Even in the absence of spin-lattice relaxation the concept "long-lived state" becomes meaningful. For a system of $N$ coupled spins, there are about $2^N$ exact integrals of motion (operators) in a total operator space of dimensionality $2^{2N}$. Projection of the density matrix onto this subspace of the integrals of motion is conserved. The resulting non-ergodic dynamics in relatively small spin clusters has been predicted by computer simulations [7,8] and experimentally confirmed [9]. Integrals of motion in small clusters of dipolar-coupled spins have been studied in [10], where it has been also noticed that the role of these integrals decreases with increasing size of the clusters. In the thermodynamic limit $N \to \infty$, a relative size of the subspace of integrals of motion becomes negligible, and the exact integrals play no role in time evolution of macroscopic observables. Only additive integrals of motion, associated with global symmetries, such as the total energy and, in our case, projection of the total spin on external magnetic field, impose explicit limitations on spin dynamics. As a consequence, the two-temperature thermodynamic theory [11] has been very successful in explaining many phenomena in solid-state NMR [12]. Rapid thermalization of a nuclear spin system is also expected from a classical picture, where each spin precesses in almost randomly fluctuating local field created by moving neighbor spins.

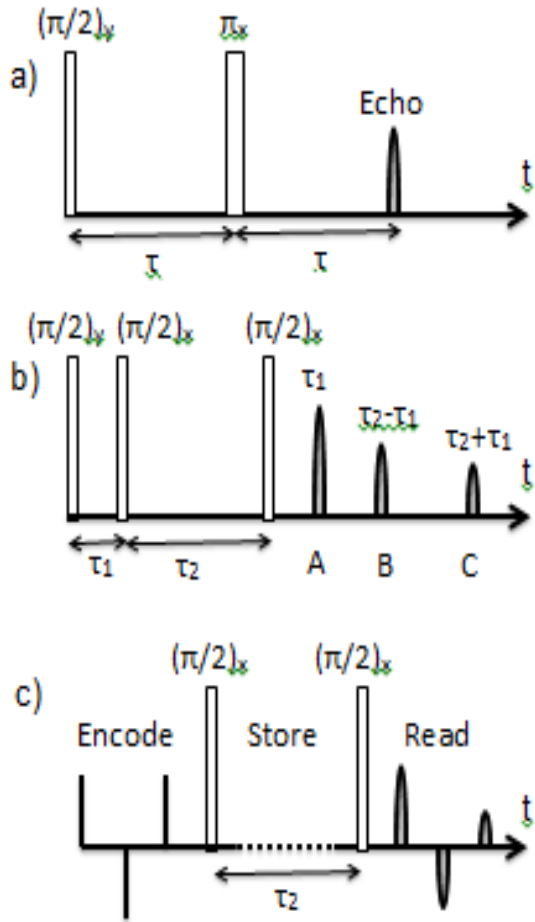

FIG. 1. Pulse sequences for exciting (a) long-lived partial echo, (b) suspended long-lived echo, and (c) information storage.

In a striking contrast to this expectation, it has been found that a weak and long radio-frequency pulse can excite a long lived response signal in many solids [13]. The signals may live orders of magnitude longer than the spin-spin relaxation time $T_2$. More recently, it became clear [14] that the phenomenon originates from the existence of a long lived partial echo in a simple spin-echo experiment (FIG.1a). Echo signals for different values of $\tau$ are shown in FIG.2 for a sample of adamantane (99+%, Aldrich). Adamantane is a plastic crystal with near-spherical fast-rotating molecules. Molecular rotations average intra-molecular dipolar couplings and chemical shift anisotropy. Each molecule contains four CH and twelve $CH_2$ protons with the difference of isotropic chemical shifts 0.1 ppm or 50 Hz in our experiments (from solution NMR). As a function of the echo time $2\tau$ the echo amplitude decays exponentially with the decay time $T_e = 12$ ms. This time is 500 times longer than the characteristic time of spin-spin relaxation $T_2 = 23$ μs. The echo amplitude, extrapolated to $\tau = 0$, $A_e = 3 \cdot 10^{-3}$. The echo width (duration), measured at ½ of its amplitude, $t_w = 550$ μs is ten times longer than $2T_2$. The echo parameters for adamantane and other samples are listed in Table 1.

**Table 1.** Echo and relaxation parameters for various samples.

|  | $T_1$, s | $T_2$, μs | $T_e$, ms | $T_{se}$, ms | $T_{se1}$, ms | $T_{se2}$, ms | $t_w$, μs |
|---|---|---|---|---|---|---|---|
| Adamantane | 1.2 | 23 | 12 | 83 | 20 | 550 | 550 |
| Naphthalene | 1700 | 4.1 | 15 | 500 | 14 | 185000 | 300 |
| Fluoroparaffin ($^1$H) | 1.1 | 7.6 | 3.5 | 29 | 8.6 | 160 | 85 |
| Fluoroparaffin ($^{19}$F) | 2.2 | 3.8 | 2.5 | 41 | 2.0 | 42 | 20 |
| 5CB | 0.95 | 8.8 | 63 | - | - | 220 | 67 |

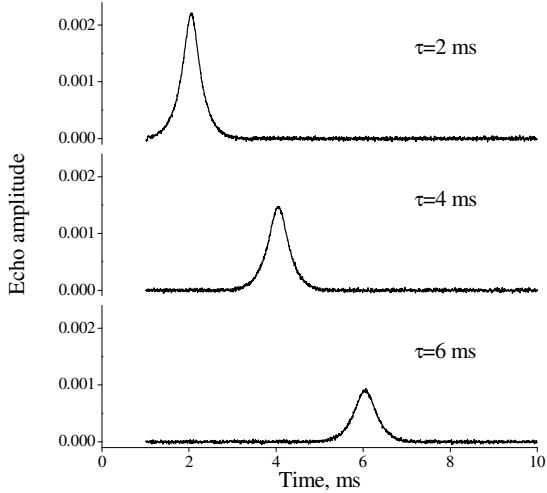

FIG. 2. Long lived echo in adamantane at different values of τ. Single-transient signals are recorded with the pulse sequence in FIG. 1a.

The echo amplitude and the decay time can be estimated as [14]

$$A_e \approx \pi\delta/\Delta\nu, \qquad T_e \approx 1/(\pi\delta), \qquad (1)$$

where δ is the distribution width for chemical shifts and Δν is the full width at half height of the conventional NMR spectrum (in Hz). For adamantane, δ = 22 Hz if calculated as a square root of the second moment. By using this value in Eqs.(1), one obtains $A_e = 5.0 \cdot 10^{-3}$ and $T_e = 14$ ms, which are not far from the experimental values. For all samples we defined $T_2$ as $T_2 = 1/(\pi\Delta\nu)$. By combining this with Eqs.(1) one obtains

$$A_e T_e / T_2 \approx \pi. \qquad (2)$$

We expect that this combination will be approximately the same for all samples, i.e. longer echo decay time means smaller echo amplitude and vice versa. Experimental values for adamantane give $A_e T_e / T_2 = 1.5$.

Even more surprising results are obtained when one uses the stimulated echo pulse sequence in FIG.1b. Similar to the case of non-interacting spins with distributed resonance frequencies (inhomogeneous broadening) [15], three echoes are observed for solids at $\tau_2 > \tau_1$: at times $\tau_1$ (A), $\tau_2 - \tau_1$ (B), and $\tau_2 + \tau_1$ (C) after the third pulse. As a function of the echo time, the echoes B and C decay similar to the echo in experiment in FIG.1a with the same decay time $T_e$. The echo A behaves differently (FIG.3). To emphasize that the mechanism of this long-lived partial echo in solids is different from that of the stimulated echo in inhomogeneously broadened systems, we will call this echo a "suspended echo". As one can see in FIG.3, the decay of the suspended echo is strongly non-exponential, so we used three different times for its characterization (Table1): $T_{se}$, the time of decaying $e$ times, $T_{se1}$, the decay time at shorter times, and $T_{se2}$, the time of exponential decay at very long times. For all the samples we studied, $T_{se}$ is significantly longer than $T_e$, $T_{se1}$ is close to $T_e$, and $T_{se2}$ is an extremely long time, comparable to $T_1$. In the suspended echo experiment, the echo amplitude depends differently on $\tau_1$ and $\tau_2$. We report dependencies on the total echo time at fixed $\tau_1$.

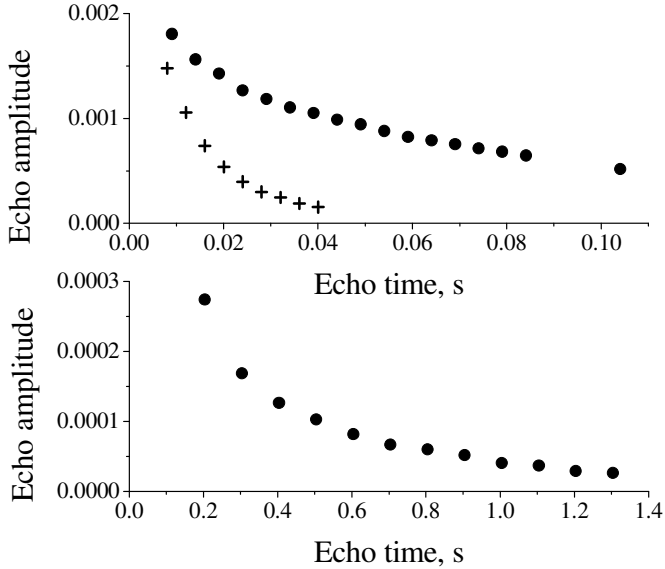

FIG. 3. Amplitude of suspended echo in adamantane as a function of the echo time at $\tau_1$ = 5 ms and varying $\tau_2$ (circles). Single-transient signals are recorded with the pulse sequence in FIG. 1b. Long-lived echo in experiment in FIG.1a is shown for comparison (crosses).

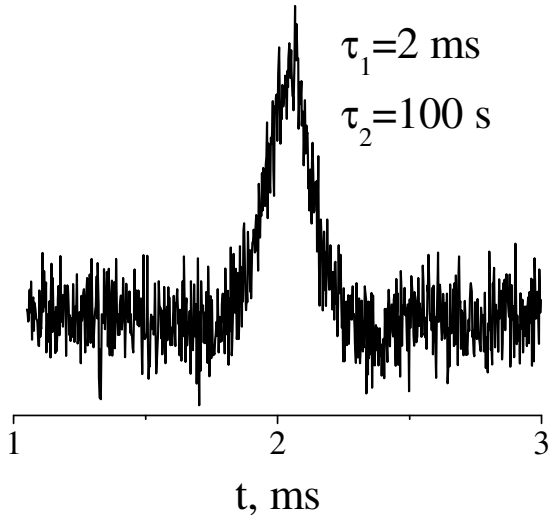

FIG. 4. Suspended echo in naphthalene, recorded with the pulse sequence in FIG.1b using $\tau_1$ = 2 ms and $\tau_2$ = 100 s. The signal is an average of 200 transients.

To illustrate that the spin state, created by the first two pulses of the pulse sequence in FIG.1b, may have a very long lifetime, we presented in FIG.4 an echo in naphthalene, recovered after 100 s "suspension time" $\tau_2$, which is seven orders of magnitude longer than the spin-spin relaxation time $T_2$. The long lived state created by the first two pulses in the suspended echo experiment "remembers" the dynamical information: the amplitude and phase of the first pulse and the time interval between the first two pulses. To further illustrate the complexity of this state, we performed the experiment in FIG1c, where the state is created by a sequence of pulses with small flip angles. The echo shape, recovered after long suspension time, reproduces this pulse sequence (in reverse time order). The result of such experiment is shown in FIG.5.

All NMR experiments have been performed with Agilent 500 MHz spectrometer using liquid-state probes. We have chosen four samples with very different structures to demonstrate the phenomenon. In adamantane, intra-molecular dipolar interactions are averaged out. In nematic liquid crystal 4'-pentyl-4-biphenyl-carbonitrile (5CB) (98%, Aldrich), inter-molecular interactions are averaged out, and the sample is an ensemble of isolated 19-spin clusters. Naphthalene (99.6%, EMD Merck) is a rigid crystal with very long spin-lattice relaxation time ($T_1$ = 1700 s). In fluorinated paraffin $CF_3(CF_2)_{11}(CH_2)_{17}CH_3$ [16], fluorination increases the range of proton chemical shifts. In addition, it is possible to make $^{19}F$ measurements. As it is expected from Eq.(1), broader range of fluorine chemical shifts leads to shorter echo times. Finite number of coupled spins in 5CB results in long components in a conventional free-induction signal. This did not allow accurate measurement of the suspended echo parameters. The echoes and relaxation characteristics for all the samples are given in Table 1. As a control sample, we used hexamethylbenzene (99+%, Aldrich), a plastic crystal with fast rotating molecules and equivalent protons. As expected, no long lived echoes have been detected for this sample.

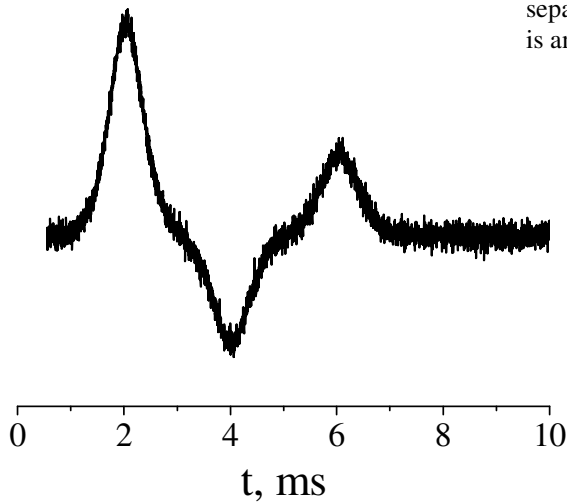

FIG. 5. Suspended echo in adamantane, recorded with the pulse sequence in FIG.1c using suspension time $\tau_2$ = 0.5 s. Three excitation pulses, as in FIG.1c, are separated by 2 ms and have 20° flip angles. The signal is an average of 4000 transients.

One can notice that, for all samples, the relations between different time scales (except $T_1$) are similar. The echo widths $t_w$ are longer than $2T_2$, but still, relatively short. This tells that the echoes formation is governed by dipole-dipole interactions, because, in our systems, there are no other interactions with appropriate strength. The difference of the echo length and shape from that of the free induction signal also tells that the spin state with transverse magnetization at the center of the echo is very different from the initial spin state created by the first excitation pulse.

In conclusion, we have demonstrated that in macroscopic systems of dipolar-coupled nuclear spins one can create spin states with extremely long life times. The states have complex structure, which "imprints" the dynamical information on how the long lived state has been created. In multi-spin systems, the echoes based on time-reversed evolution have relatively short lifetime. As an example the "magic echo" [17] lives about 10 $T_2$. When effective Hamiltonians are used for both forward and backward evolution [18], the echo lifetime can be increased by one more order of magnitude. As we demonstrated here, small-amplitude echo in a stimulated echo experiment in FIG.1b can live seven orders of magnitude longer than $T_2$.

The authors thank Drs. A. Jerschow and J.-S. Lee for stimulating discussions, G. Bodenhausen for pointing at analogies between the long-lived echo in solids and long-lived coherences in spin pairs, and R.J. Twieg for providing a sample of fluoroparaffin. We also acknowledge a support from NSF CHE-1048645 (AK) and the Fulbright Foundation (AT).